\newcommand\arxivversion[2]{#1}  

\arxivversion{
  \documentclass[twocolumn,10pt]{article}
}{
  \documentclass[
    ,final            
  ]
                {aipproc}
}

\usepackage{amsfonts}

\arxivversion{
  \usepackage{natbib}
  \usepackage{graphicx}
  \newenvironment{theacknowledgments}{\section*{Acknowledgments}}{}
  \usepackage{fullpage}
}{
  \relax
}

\usepackage{hyperref}
\newcommand\eprint[2][]{\href{http://arXiv.org/abs/#2}{{\tt #2}}}

\newcommand\md{\mathrm{d}} 
\newcommand\me{\mathrm{e}} 

\providecommand\apj{ApJ}                 
\providecommand\cqg{ClassQuantGrav}                       
\providecommand\aap{A\&A}            
\providecommand\mnras{MNRAS}
\providecommand\BASI{BullAstrSocIndia}

\providecommand\prd{Phys. Rev. D.}
\providecommand\PRL{Phys. Rev. Lett.}

\providecommand\physrep{Physics Reports}
\providecommand\nat{Nature}

\newcommand\hMpc{\mbox{$h^{-1}$ Mpc}}
\newcommand\hGpc{\mbox{$h^{-1}$ Gpc}}

\newcommand\gtapprox{\,\lower.6ex\hbox{$\buildrel >\over \sim$} \, }
\newcommand\ltapprox{\,\lower.6ex\hbox{$\buildrel <\over \sim$} \, }

\arxivversion{}{
  \layoutstyle{8x11double}
}

\begin{document}

\title{Topological acceleration in relativistic cosmology}

\arxivversion{
  \date{
  }
}{
  \protect\classification{98.80.-k, 98.80.Es, 98.80.Jk}

  \protect\keywords      {cosmology: observations -- cosmology: theory -- cosmic microwave background}
}

\arxivversion{
  \protect\author{Boudewijn F. Roukema\\
    Toru\'n Centre for Astronomy, Nicolaus Copernicus University,
    ul. Gagarina 11\\ 87-100 Toru\'n, Poland}
}{
  \protect\author{Boudewijn F. Roukema}{
    address={Toru\'n Centre for Astronomy, Nicolaus Copernicus University,
      ul. Gagarina 11, 87-100 Toru\'n, Poland}}
}

\arxivversion{
  \maketitle
}{
}

\begin{abstract}
Heuristic approaches in cosmology bypass more difficult calculations
that would more strictly agree with the standard Einstein
equation. These give us the well-known
Friedmann-Lemaitre-Robertson-Walker (FLRW) models, and, more recently,
the feedback effect of the global topology of spatial sections on the
acceleration of test particles. Forcing the FLRW heuristic model on
observations leads to dark energy, which, pending fully relativistic
calculations, is best interpreted as an artefact. Could topological
acceleration also be an artefact of using a heuristic approach? 
A multiply connected exact solution of the Einstein equation shows
that topological acceleration is present in at least one fully relativistic
case---it is not an artefact of Newtonian-like thinking.
\end{abstract}

\arxivversion{
}{
  \maketitle
}


\section{Can we go beyond heuristics?}
A spatial section of the present-day Universe is clearly
inhomogeneous, i.e. the density $\rho(\mathbf{x},t)$ over the spatial
section (3-manifold) at a given cosmological time $t$ cannot be
written as a function of $t$ alone. But finding a general,
inhomogeneous solution to the Einstein field equation is
difficult. This is why the standard family of cosmological models
consists of the Friedmann-Lemaitre-Robertson-Walker (FLRW) models with
constant curvature at any given cosmological time $t$, i.e.
$\rho(\mathbf{x},t) = \rho(t),$ and a fixed 3-manifold topology
consistent with the comoving curvature
\citep{deSitt17,Fried23,Fried24,Lemaitre31ell,Rob35}.  Strictly
speaking, the models are inconsistent with the existence of
galaxies. This is bypassed by the heuristic approach of
(mathematically) hypothesising that perturbations on the exact
solution can be evolved in such a way that the perturbed global
spacetime remains a solution of the Einstein equation. The Concordance
Model estimates of the FLRW model parameters \citep[][, refs therein,
  and citations thereof]{CosConcord95} show that forcing the FLRW
model on extragalactic observations requires the existence of a
cosmological constant or dark energy parameter $\Omega_\Lambda$, which
becomes significant at approximately the epoch when overdensities
become significantly non-linear. Work towards more realistic, physical
models indicates that the simplest interpretation of $\Omega_\Lambda$
is that it is an artefact of forcing an oversimplified model (the FLRW
model) onto real world data
\citep[e.g.][]{CBK10,BuchCarf03,WiegBuch10}.

 \begin{figure*}
   \includegraphics[width=0.9\textwidth]{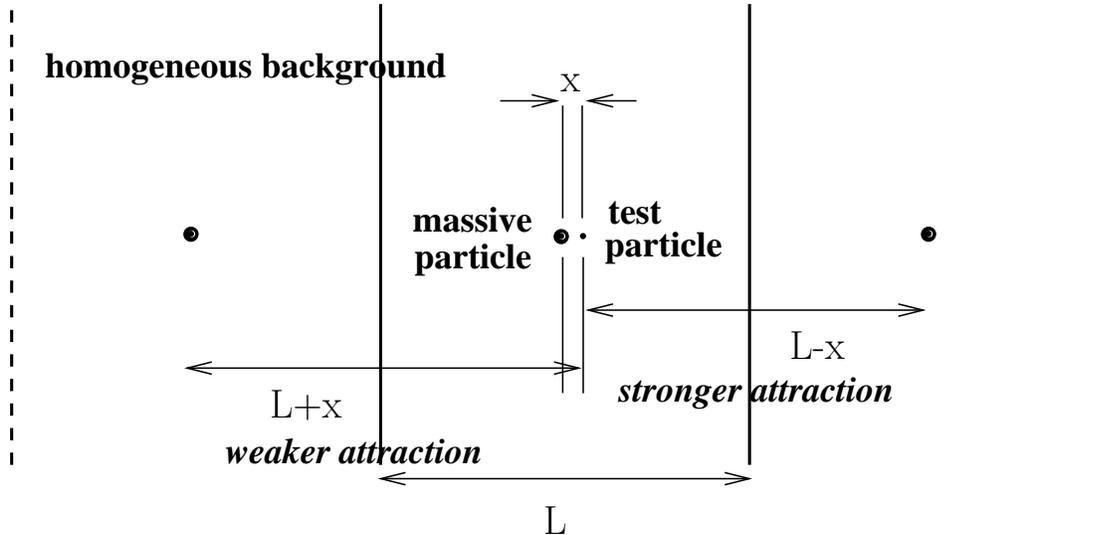}
   \caption{Two-dimensional cut through a $T^1$ (slab space) model,
     showing the ``zeroth'' copy of the massive object and adjacent
     copies in the universal covering space at left and right,
     separated from the first copy by the fundamental domain length
     $L$. If the third spatial dimension is ignored, this can be
     thought of as an ordinary cylinder.  A massless test particle at
     distance $x$ from the zeroth copy of the massive particle is
     shown.
 \label{f-T1}}
 \end{figure*}

A complementary heuristic calculation is the effect of the topology
of the spatial section 
(\citep{LaLu95,Lum98,Stark98,LR99,RG04}; or shorter:
\citep{Rouk00BASI}) 
on the dynamics. Again, in a strictly FLRW
model, there is no effect of topology on dynamics (apart from 
the link between the sign of curvature and the family of constant
curvature 3-manifolds). Since the real Universe is inhomogeneous,
the first way of determining the nature of the effect
of topology on dynamics was heuristical. The acceleration of 
a test particle towards a single point-like massive object in 
a homogeneous background was calculated in \citet{RBBSJ06} for
a slab space (one multiply-connected direction with zero curvature, $T^1$)
and a 3-torus ($T^3$) model. Topological acceleration was found
to exist at linear order in the (small) displacement of the test
particle from the massive particle for the $T^1$ case and the
irregular $T^3$ case (unequal fundamental lengths). In a slab space of 
injectivity diameter (``box size'') $L$,
if we place the test particle along the line joining the multiple images
of the massive object of mass $M$, at a small separation $x$ from a copy of the massive
object, then the test particle is at $L-x$ from one ``adjacent distant'' copy
of the massive particle (in the universal covering space)  and at 
$L+x$ from the other ``adjacent distant'' copy
of the massive particle (see Fig.~\ref{f-T1}). Thus, the pseudo-Newtonian 
topological acceleration (ignoring the attraction towards the close copy of the
massive object) is \citep{RBBSJ06}
\begin{eqnarray}
  \ddot{x}_{\mathrm{topo}} &=&
  M \left[ \frac{1}{(L-x)^2} - \frac{1}{(L+x)^2}  \right]
  \nonumber \\
  &\approx &  \frac{4M}{L^2}\;\frac{x}{L}  
  \label{e-topaccel-slab-newt}
\end{eqnarray}
to first order in $x/L \ll 1$, where the gravitational constant
is in natural units ($G=1$). 
Inclusion of more distant copies of the massive object without
bound gives a slightly higher value:
\begin{eqnarray}
  \ddot{x}_{\mathrm{topo}} &=& M \sum_{j=1}^{\infty} 
  \left[ \frac{1}{(jL-x)^2} - \frac{1}{(jL+x)^2}  \right] \nonumber \\
  &=&  4\zeta(3)  \frac{M}{L^2} \;\frac{x}{L}   
  \label{e-topaccel-slab-newt-inf}
\end{eqnarray}
using Ap\'ery's constant $\zeta(3)$, where $\zeta$ is the
Riemann zeta function.

Curiously, topological acceleration, calculated according to this
heuristical approach, cancels to first order in a regular $T^3$ model
\citep{RBBSJ06}.  In the three spherical 3-manifolds with the most
symmetrical properties, i.e. the ``well-proportioned'' ones
\citep{WeeksWellProp04}, topological acceleration also cancels to
first order \citep{RR09}. In two of them, the octahedral space
$S^3/T^*$ and the truncated-cube space $S^3/O^*$ \citep{GausSph01},
the effect reappears at third order, as it does for the regular $T^3$
model. In the third space, the Poincar\'e dodecahedral space
$S^3/I^*$, the cubical term also cancels exactly, leaving a fifth
order term. Thus, topological acceleration not only shows that global
topology has an effect on the dynamics of a universe, but that
it distinguishes different topologies. The particular way that it
distinguishes different 3-manifolds happens to select the Poincar\'e
space as being special---not just ``well balanced'', but ``very well
balanced''---independently of the cosmic microwave background
observational arguments first presented
several years earlier in favour of the Poincar\'e space
\citep{LumNat03,Aurich2005a,Aurich2005b,Gundermann2005,Caillerie07,RBSG08,RBG08}.

Yet, just as we lack evidence for ``dark energy'' (and acceleration of the
scale factor $a(t)$) being anything more
than an artefact of a heuristical approach to solving the Einstein
equation, could the topological acceleration calculations
\citep{RBBSJ06,RR09} also be an artefact of a heuristical, i.e.
pseudo-Newtonian approach?

\section{Compact Schwarzschild spacetime}
An exact Schwarzschild-like solution of the Einstein equations is
known for which the spatial topology (apart from the ``central'' black hole
singularity itself) is the slab space \citep{KN94}. For a test
particle distant from the event horizon and an injectivity diameter ($L$)
which is much greater still, it should be possible to check whether
Eq.~(\ref{e-topaccel-slab-newt-inf}) is a limiting case of this exact,
``compact Schwarzschild'' solution. 

This was checked in \citet{ORB11}. 
Using the conventions $G=c=1$,
a signature of $(-,+,+,+)$, Greek spacetime indices $0,1,2,3$ 
and Roman space indices $1,2,3$, the following assumptions
can be made \citep{ORB11}.
The test particle is distant from the event horizon but much closer to the
``zeroth'' copy of the massive object than to any other copies
in the universal covering space:
\begin{equation} 
  0 < M \ll x \ll L/2 ;
  \label{e-x-assump}
\end{equation}
the test particle has a low coordinate velocity
(implying a low 4-velocity spatial component):
\begin{equation} 
  \frac{\mathrm{d}x^i }{ \mathrm{d}t} \ll 1 
\;\;
\Rightarrow \;\;
  \frac{\mathrm{d}x^i }{ \mathrm{d}\tau} \ll 
  1 \ltapprox
  \frac{\mathrm{d}t}{\mathrm{d}\tau } ;
  \label{e-low-beta}
\end{equation}
and the spacetime is a vacuum model:
\begin{equation} 
  \rho = 0 ,
  \label{e-zero-rho}
\end{equation}
where the proper time $\tau$ along the test particle's world line 
parametrises the latter, i.e. $x^\alpha(\tau)$, and $t \equiv x^0$. 

Following \citet{KN94},
the metric is given in Weyl coordinates, using the Ernst potential:
\begin{equation}
  \md s^2 = -\me^{\omega} \md t^2 +
  \me^{-\omega} \left[ \me^{2k} (\md x^2 + \md \rho^2) + \rho^2 \md \phi^2 \right]
  \label{e-KN-metric}
\end{equation}
from Eq.~(9) of \citet{KN94},  where $k$ and $\omega$ are related to an
Ernst potential $\varepsilon$ defined on $\xi := x + \mathrm{i} \rho$,
and
\begin{equation}
  \varepsilon_0(x,\rho) := 
  \frac{ \sqrt{(x-M)^2 +\rho^2} + \sqrt{(x+M)^2 +\rho^2} -2M}{
   \sqrt{(x-M)^2 +\rho^2} + \sqrt{(x+M)^2 +\rho^2} +2M}
  \label{e-defn-vareps_0} 
\end{equation}
\begin{equation}
  \omega_0 := \ln \varepsilon_0,\; a_0 := 0,\; 
  {{{a_{j\not=0} := \frac{2M}{L\,|j|}}}}
  \label{e-defn-omega_0}
\end{equation}
\begin{equation}
  \omega(x,\rho) := \sum_{j=-\infty}^{\infty} \left[\omega_0(x+jL,\rho) + a_j\right]
  \label{e-defn-omega}
  ,\;\;\;
  \varepsilon := \me^\omega
  \label{e-defn-vareps}
\end{equation}
\begin{equation}
  \partial_\xi k = 2 \mathrm{i} \rho 
  \frac{\partial_\xi \varepsilon \partial_\xi \bar{\varepsilon}}{
    (\varepsilon + \bar{\varepsilon})^2},
  \label{e-defn-k}
\end{equation}
i.e. Eqs~(13), (12), (5) and (7) of \citet{KN94}.
As shown in \citet{ORB11}, 
the topological acceleration of a test particle under these 
conditions is identical to that of Eq.~(\ref{e-topaccel-slab-newt-inf})

Numerical evaluation of $\ddot{x}_{\mathrm{topo}}$ using the Weyl
metric and 100-bit arithmetic 
shows that for $M \sim 10^{14} M_\odot$, $L \sim 10$ to $20{\hGpc}$, 
the linear expression is quite accurate over several orders of 
magnitude in length scale, i.e. the linear expression and the
numerical estimate agree to within $\pm10\%$ 
for $3{\hMpc} \ltapprox x \ltapprox  2{\hGpc}$ \citep{ORB11}.

\section{Conclusion}
Thus, it is clear that topological acceleration is not a Newtonian
artefact: it exists in at least one relativistic spacetime, and most
likely in others.  Moreover, given some typical astronomical values
for the parameters of the model, the linear (Newtonian-like,
first-order) approximation of the effect is a good estimate of the
effect over several orders of magnitude in length scale. 
It remains
to be seen if exact solutions can be found for $T^3$, $S^3/T^*$,
$S^3/O^*$ and $S^3/I^*$ spatial sections, and if the linear terms
cancel exactly as in the pseudo-Newtonian case.

\begin{theacknowledgments}
A part of this project has made use of 
Program Oblicze\'n Wielkich Wyzwa\'n nauki i techniki (POWIEW)
computational resources (grant 87) at the Pozna\'n 
Supercomputing and Networking Center (PCSS).
\end{theacknowledgments}



\bibliographystyle{aipproc}   




\IfFileExists{\jobname.bbl}{}
 {\typeout{}
  \typeout{******************************************}
  \typeout{** Please run "bibtex \jobname" to optain}
  \typeout{** the bibliography and then re-run LaTeX}
  \typeout{** twice to fix the references!}
  \typeout{******************************************}
  \typeout{}
 }

\end{document}